\documentstyle[epsfig,epsf]{article}
\def\Dm {\Delta m^2}
\def\stheta {\sin^2 2\theta}

\newcommand{\nm}{\rm \,nm}

\def\ne {\nu_e}
\def\nm {\nu_\mu}
\def\nt {\nu_\tau}

\def\nmne {\nu_\mu \longleftrightarrow \nu_e}
\def\nmnt {\nu_\mu \longleftrightarrow \nu_\tau}
\def\nmns {\nu_\mu \longleftrightarrow \nu_{sterile}}

\begin{document}

\begin{center}
{\Large {\bf Atmospheric {\boldmath $\nu$} and Long Baseline {\boldmath $\nu$} 
experiments}}
\end{center}

\vskip .7 cm

\begin{center}
Giorgio Giacomelli 
 \par~\par
{\it Dept of Physics, Univ. of Bologna, and INFN, \\
V.le C. Berti Pichat 6/2, Bologna, I-40127, Italy\\} 

E-mail: giacomelli@bo.infn.it

\par~\par

Invited Lecture at the Carpatian Summer School of Physics 2007, Sinaia, 
Romania, August 2007

\vskip .7 cm
{\large \bf Abstract}\par
\end{center}

{\normalsize 
The results obtained by several experiments on atmospheric neutrino 
oscillations are summarized and discussed. Then the results obtained by 
different long baseline neutrino experiments are considered. Finally 
conclusions and perspectives are made.}


\section{Introduction}\label{sec:intro}

Atmospheric neutrinos are well suited for 
the study
of neutrino oscillations, since they have energies from a fraction of GeV up 
to more than 100 GeV and they may travel distances $L$ from few tens of km 
up to 13000 km; thus $L/E_\nu$ ranges from $\sim$1 km/GeV to $10^5$ 
km/GeV. Atmospheric neutrinos may study $\nu$ oscillations
for small $\Delta m^2$. \par

The early water Cherenkov detectors IMB \cite{imb}
and Kamiokande \cite{kamioka} reported 
anomalies in the ratio of muon to electron
neutrinos, while the tracking calorimeters NUSEX \cite{nusex} and 
 Frejus \cite{frejus}, and the Baksan \cite{Baksan} scintillator detector
did not find any. In 1995 MACRO found a deficit for upthrougoing muons 
\cite{mac17}. Then the Soudan 2 experiment \cite{soud2} confirmed the ratio 
anomaly. In 1998 Soudan 2 \cite{soud2_b}, MACRO \cite{macro98} and 
SuperKamiokande (SK) \cite{skam} reported deficits in the $\nu_\mu$ 
fluxes with
respect to Monte Carlo (MC) predictions and angular distribution 
distortions; instead the $\nu_e$ distributions were in agreement with 
non oscillated MCs. These features
may be explained in terms of $\nu_\mu \longleftrightarrow
\nu_\tau$ oscillations. \par

The atmospheric neutrino flux was computed by many authors in the mid 1990s 
\cite{Bartol96} and in the early 2000s \cite{honda01}. The last had 
improvements, but also a new scale uncertainty. \par

Several long baseline $\nu$ beams were and are operational: KEK to Kamioka 
(K2K) (250 km), NuMi from Fermilab to the Soudan mine (730 km) and CERN 
to Gran Sasso(GS) (CNGS) (730 km). The K2K \cite{k2k} and MINOS \cite{minos} 
experiments obtained results in agreement with the atmospheric $\nu$ 
results. The CNGS beam is under tests: some events were recorded in OPERA 
\cite{opera} and LVD \cite{lvd} at GS. \par
Most experiments are disappearance experiments; OPERA is an appearance 
experiment, searching for $\nu_{\tau}$ in a pure $\nu_{\mu}$ beam. \par
Long baseline experiments will give an increasing contribution to neutrino 
physics.

  \section{Atmospheric neutrino oscillations}
A high energy primary cosmic ray (CR), proton or nucleus, interacts in the upper atmosphere producing  a large number of charged pions and kaons, which decay yielding  muons and $\nu_{\mu}$'s; also the muons decay yielding \( \nu _{\mu } \) and \( \nu _{e } \). The ratio of the numbers of \( \nu _{\mu } \) to \( \nu _{e } \)  is $\simeq$  2 
and $N_{\nu}/N_{\overline\nu} \simeq 1$. Atmospheric  neutrinos are produced at  10-20 km above ground, and they proceed towards the earth. 

If $\nu$'s have non-zero masses, one considers the {\it weak flavour eigenstates} $~\ne,~\nm,~\nt$ and the {\it mass eigenstates} $~\nu_1,~\nu_2,~\nu_3$.  Flavour eigenstates are linear combinations of  mass eigenstates. For 2 flavour  ($\nm,~\nt$) and 2 mass eigenstates $(\nu_2,~\nu_3)$ one writes
 
\begin{equation}
\left\{ \begin{array}{ll}
      \nm =~\nu_2 \cos\ \theta_{23} + \nu_3 \sin\ \theta_{23} \\
      \nt=-\nu_2\sin\ \theta_{23} + \nu_3\cos\ \theta_{23} 
\end{array} 
\right.
\end{equation} 
\noindent where $\theta_{23}$ is the mixing angle. The survival probability of the $\nm$ ``beam" is
\begin{equation}
P(\nm \rightarrow \nm) = 1- P(\nm \rightarrow \nt) = 1- \sin^2 2\theta_{23}~\sin^2 \left( 
{\frac{1.27 \Dm \cdot L} {E_\nu}} \right)
\label{eq:prob}
\end{equation}
\noindent where $\Dm=m^2_3-m^2_2$ and $L$ is the distance from $\nu$ production to detection. 

The simple formula Eq. \ref{eq:prob} is modified by additional flavours and by matter effects.

\section{Results from atmospheric neutrino experiments}
After the early indications of an anomaly in atmospheric $\nu$'s [1-7], in 1998 Soudan 2, MACRO and SK provided strong indications in favour of $\nmnt$ oscillations  [8-10] \cite{macro}.  Confirming results were presented by long baseline experiments \cite{k2k} \cite{minos}. \par

{\bf Soudan 2} used a modular fine grained tracking and showering
calorimeter of $963$ t located underground in the
Soudan Gold mine in Minnesota. The detector 
was made of 1m$\times$1m$\times$2.5m modules, surrounded by an anticoincidence  shield. The
bulk of the mass was 1.6 mm thick corrugated  steel sheets 
 interleaved with drift tubes \cite{soud2_b}. The Soudan 2 
double ratio for the zenith angle range $-1 \leq \cos \Theta \leq 1$ 
is $R^\prime=(N_\mu/ N_e)_{DATA}/ (N_\mu/ N_e)_{MC} 
= 0.68 \pm 0.11_{stat} $, consistent with $\nm$ oscillations. \par

\indent {\bf MACRO} detected upgoing $\nm$'s via CC interactions $\nm \rightarrow \mu$; upgoing muons were identified with streamer tubes (for tracking) and scintillators (for time-of-flight measurements), points and rectangles in Fig. 1a \cite{tecnico}. Events were classified as:\\
{\it Upthroughgoing muons} from interactions in the rock below the detector of $\nm$ with $\langle E_\nu \rangle \sim 50$ GeV. Data were compared with the predictions of the Bartol96 \cite{Bartol96}, FLUKA, HKKM01 \cite{honda01} MCs,  Fig. \ref{fig:cosze}a. The shape of the angular distribution and the absolute value favoured $\nu$ oscillations with $\Dm = 0.0025$ eV$^2$ and maximum mixing.
The absolute value of the muons is $25\%$ higher than FLUKA and HKKM01 MC predictions . The difference between new and old MC predictions is due to new fits of CR data \cite{gaisser}. \par

{\it Low energy events. Semicontained upgoing muons} (IU) come from
$\nm$ interactions inside the lower apparatus. {\it Up stopping muons} (UGS)
are due to external $\nm$ interactions yielding upgoing muons stopping in the 
detector; the {\it semicontained downgoing muons} (ID) are due to downgoing 
$\nm$'s interacting in the lower detector; the lack
of time information prevents to distinguish between the two samples.
An equal number of UGS and ID events is expected. The average 
parent neutrino energy is 2-3 GeV. 
The data are compared with  MC
predictions without oscillations in   Figs. \ref{fig:cosze}b,c: they show a uniform deficit 
over the whole angular distribution with respect to Bartol96 predictions. \par

\begin{figure}
\begin{center}
\includegraphics[height=2.6in]{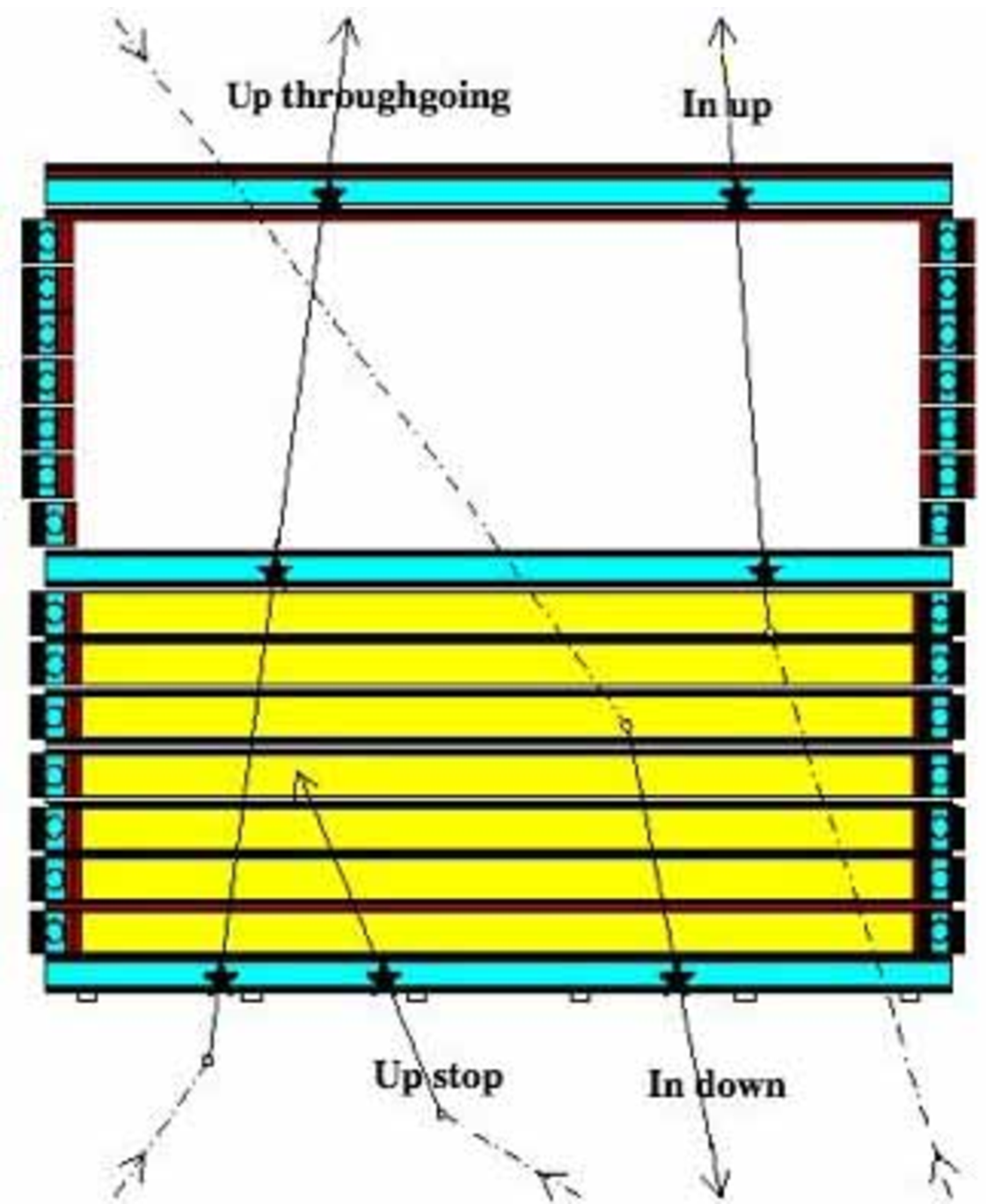}
\hspace{0.2in}
\includegraphics[height=2.2in]{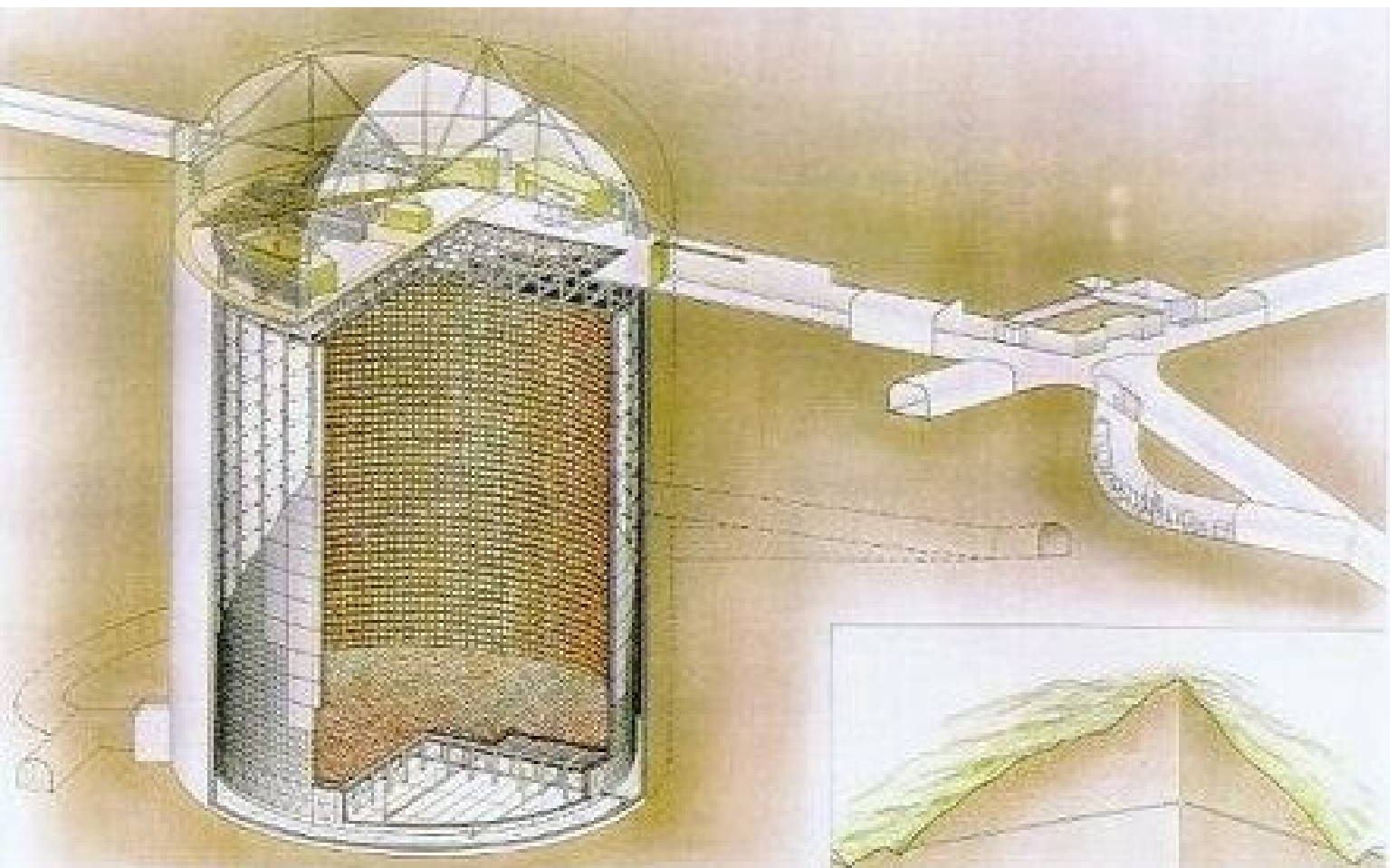}
\caption{\label{label2} a)  Cross section of the MACRO detector, sketch of event topologies. b) Schematic layout of the SuperKamiokande detector}
\end{center}
\end{figure}

{\it $\nmnt$ against $\nmns$.} Matter effects due to 
the difference between the weak interaction effective potential for 
$\nu_{\mu}$ with respect to sterile neutrinos, yield different total number and different zenith 
distributions of upthroughgoing muons. $R_{meas}$ between the events with $-1 < \cos \Theta < -0.7$ and with 
$-0.4 < \cos \Theta < 0$ was used \cite{macro98}.  The measured ratio, $R_{meas}=1.38$, is compared with $R_\tau=1.61$ and
$R_{sterile}=2.03$. One concludes that $\nmns$ oscillations are excluded at the 99.8\% c.l. \par

{\it $\nm$ energy estimate by Multiple Coulomb Scattering (MCS) of upthroughgoing muons.}
An estimate of the muon energy was made through their Multiple Coulomb 
Scattering (MCS) in the absorbers \cite{mulsca}. The ratios 
Data/MC$_{\mbox{no~osc}}$ as a function of 
$(L/E_\nu)$ are in agreement with  $\nmnt$ oscillations \cite{mac52}. \par

{\it New determination of the oscillation parameters.} In order to reduce 
the effects of systematic uncertainties in the MCs, MACRO 
 used the following three independent ratios (it was checked
 that FLUKA, HKKM01 and Bartol96 MC yield the same 
predictions):

(i) High Energy Data: zenith distribution ratio: $R_1 = N_{vert}/N_{hor}$

(ii) High Energy Data: $\nu_{\mu}$ energy measurement ratio: $R_2 = N_{low}/N_{high}$

(iii) Low Energy Data:  $R_3 = (Data/MC)_{IU}/(Data/MC)_{ID+UGS}$.

\noindent The no oscillation hypothesis had a probability 
P$\sim$3 $\cdot 10^{-7} $
and is ruled out by $ \sim 5 \sigma$. Fitting the 3 ratios 
to the $\nmnt$ oscillation formulae, one obtained $\stheta = 1,~\Dm = 2.3
\cdot 10^{-3}$ eV$^2$. Using  Bartol96, one adds the 
information on absolute fluxes:

   (iv) High energy data (systematic error $\simeq$17$\%$): $R_4 = N_{meas}/N_{MC}$.

   (v) Low energy semicontained muons (scale error $21 \%$): $R_5 = N_{meas}/N_{MC}$.

\begin{figure}
\includegraphics[height=1.8in]{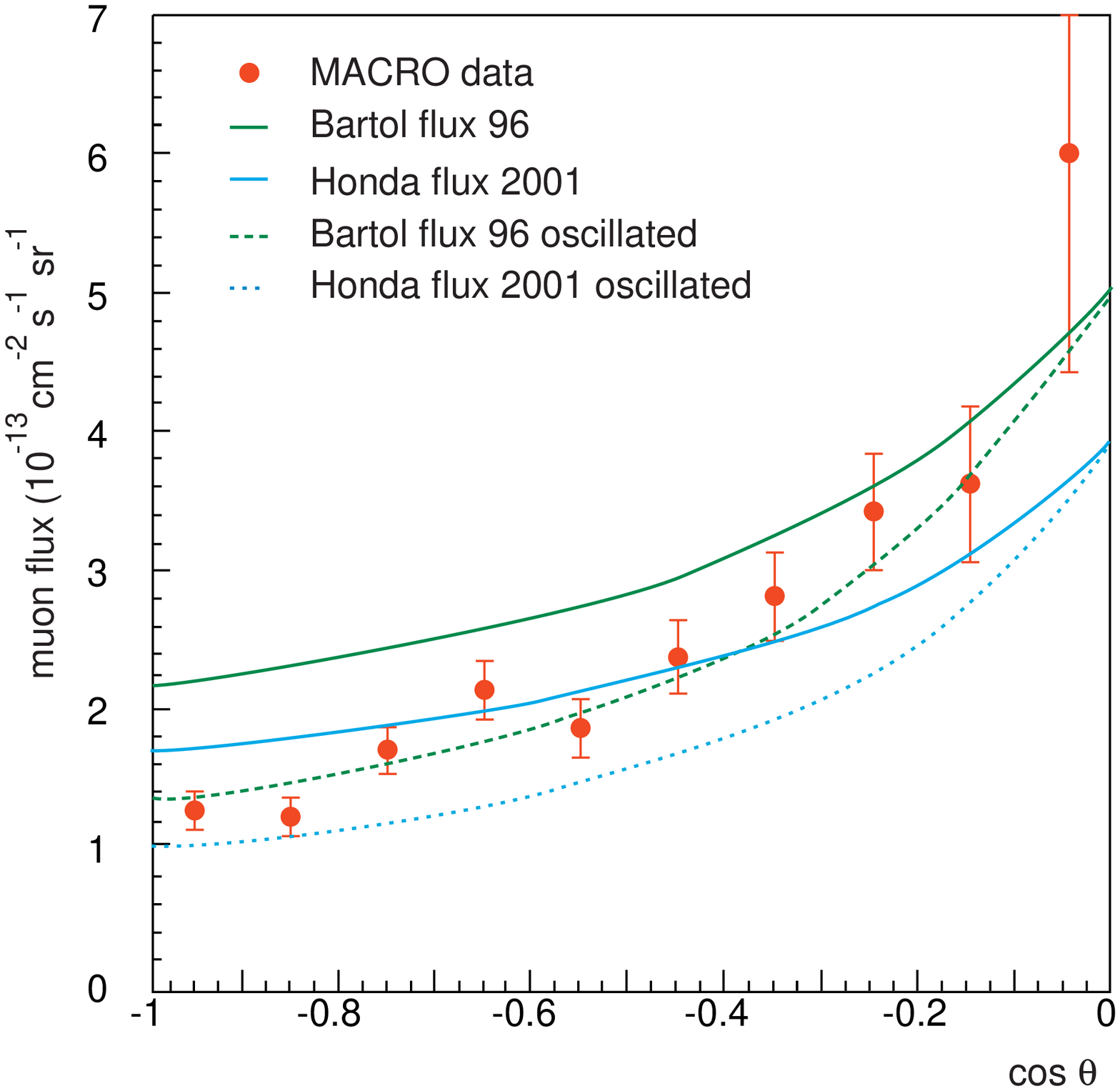}
\includegraphics[height=1.8in]{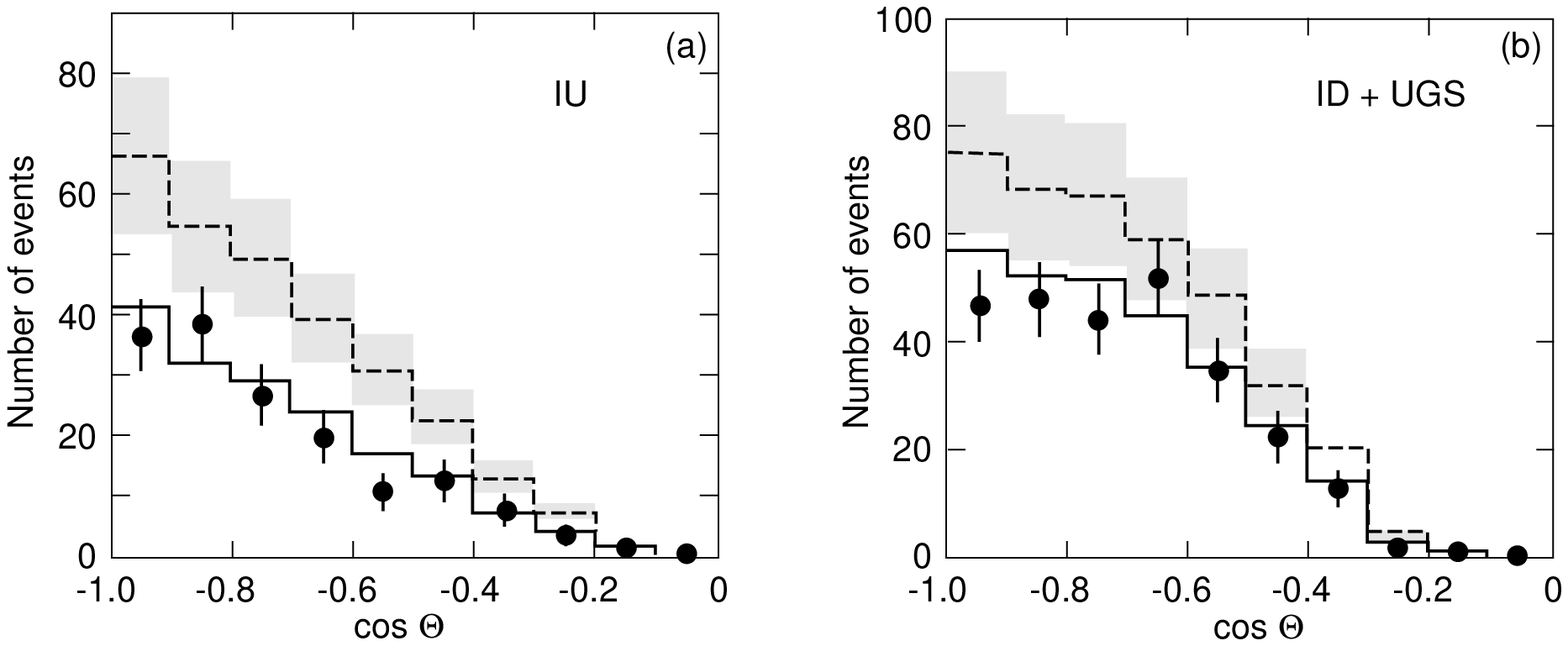}
\caption { \label{fig:cosze} {left) MACRO upthroughgoing muons compared with the oscillated MC predictions of Bartol96 (solid curve), HKKM01 (dash-dotted line), FLUKA fitted to the new CR fit (dashed curve) and FLUKA with the old CR fit (dotted curve). center) IU and  right) ID+UGS  events (black points) compared with the non oscillated Bartol96 MC (dashed line) and $\nmnt$ predictions.}}
\end{figure}

\noindent These leave the best fit values unchanged and improve the significance to $6\sigma$.\par

{\bf SuperKamiokande} (SK) is a large cylindrical water 
Cherenkov detector 
of 39 m diameter and 41 m height containing 50 kt of 
water (fiducial mass 22.5 kt); it is seen by 50-cm-diameter 
inner-facing phototubes (PMTs), Fig. 1b. The 2 m thick outer layer of water acts 
as an anticoincidence; it is seen by smaller outward-facing PMTs. The detector 
is located in the Kamioka mine, Japan. Atmospheric neutrinos are detected in SK by measuring the Cherenkov light generated by the charged particles produced in the neutrino CC interactions with water nuclei.  Thanks to the high PMT coverage, the experiment detects events of energies as low as $\sim$5 MeV.

 The large detector mass allows to collect a high statistics sample of {\it fully contained} events ({\it FC}) up to $\sim$5 GeV. The {\it FC} events yield rings of Cherenkov light on the 
PMTs. {\it FC} events can be subdivided
into {\it sub-GeV} and {\it multi-GeV} events, with energies below and above 1.33 {\it GeV}. {\it FC} events include only single-ring events,
while {\it multi-ring} ones ({\it MRING}) are treated as a separate category.
The {\it partially contained} events ({\it PC}) are CC interactions with vertex within the fiducial volume and at least a charged 
particle, typically
the $\mu$, exits the detector (the light pattern is a filled circle).  {\it Upward-going muons} ({\it UPMU}), produced by 
$\nu_{\mu}$ from below interacting in the rock, are subdivided into 
{\it stopping} ($\langle E_\nu \rangle \sim 7$ GeV) and {\it throughgoing muons} ($\langle E_\nu \rangle \sim 70 \div 80$ GeV) \cite{skam}.\par

\begin{figure}[!h]
\begin{center}
\includegraphics[width=26.5pc]{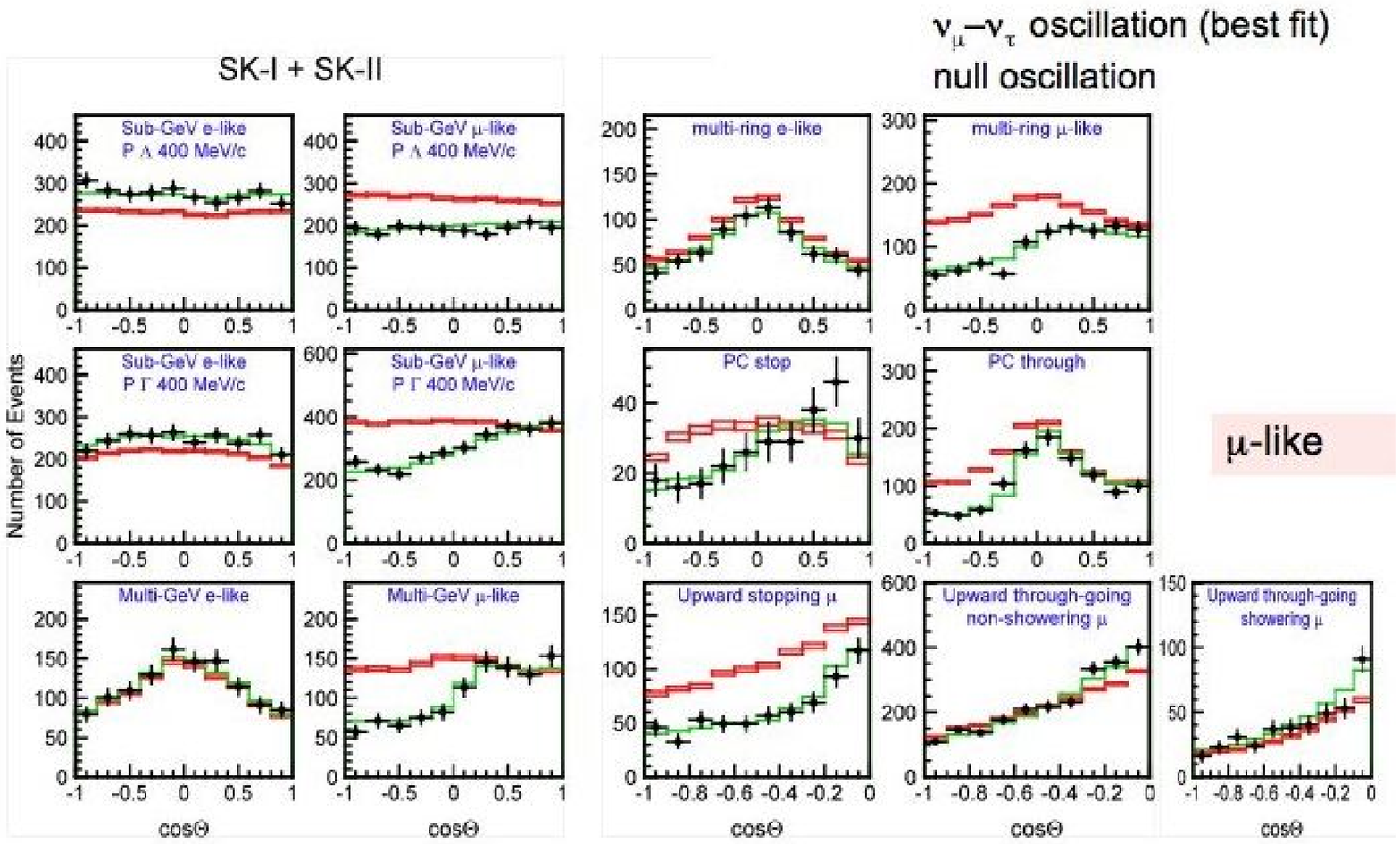}
\includegraphics[width=11pc]{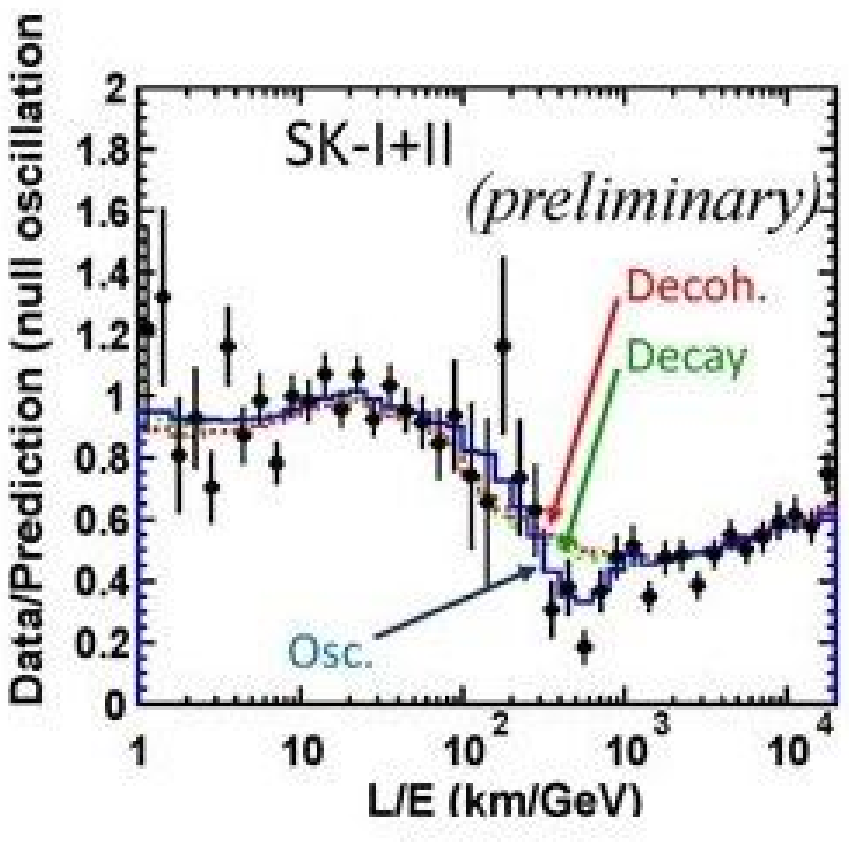}
\caption{\label{SK} Left: SK zenith distributions (black points) for $e$-like and $\mu$-like sub-GeV, multi-GeV, throughgoing and stopping muons. Solid lines are no oscillation MC predictions. Right: L/E$_{\nu}$ distribution for $\mu$-like events.} 
\end{center}
\end{figure}

Particle identification is performed using
likelihood functions to parametrize the sharpness of
the Cherenkov rings, which are more diffused for electrons than for
muons. The zenith angle distributions for $e$-like and $\mu$-like 
{\it sub-GeV} and {\it multi-GeV} events are shown in Fig. 3 left. The MC problem described before exists also in SK \cite{skam}: the e-like events were in agreement with the HKKM95 MC predictions for no-oscillations; they are higher than the HKKM01 non oscillated MC. For  $\mu$-like events, the new MC predictions are low: to reduce these problems the normalization is a free parameter.\par

The ratios $e$-like events/MC do not depend from $L/E_\nu$ 
 while $\mu$-like events/MC show a
dependence on $L/E_\nu$ consistent
with an oscillation hypothesis, Fig. 3 right. The overall best fit of SK data corresponds to maximal mixing and $\Delta m^2 = 2.5 \cdot 10^{-3}$ eV$^2$.

{\bf Exotic oscillations.} MACRO and SuperK data were used to search for sub-dominant oscillations due to Lorentz invariance violation (or violation of the equivalence principle). In the first case there could be mixing between flavor and velocity eigenstates. 90$\%$ c.l. limits were placed in the Lorentz invariance violation parameters 
$\left|\Delta v\right|<6\cdot10^{-24}$ at ${\stheta}_v$=0 and $\left|\Delta v\right|<4\cdot10^{-26}$ at ${\stheta}_v=\pm$1 \cite{lorentz}. {\bf Neutrino decay} could be another exotic explanation for neutrino disappearence; no radiative decay was observed \cite{notte}.

\section{Long baseline neutrino beams and experiments}
Neutrino physics has opened new windows into phenomena beyond the Standard 
Model of particle physics. Long baseline neutrino experiments may allow
further insight into $\nu$ physics. The first long baseline $\nu$  beam was the KEK to Kamioka (K2K) beam, the 2$^{nd}$ was the Fermilab to the Soudan mine beam (NuMi). CNGS \cite{CNGS} was commissioned in 2006 and started sending neutrinos to the GS Lab.  

\begin{figure}[!h]
\includegraphics[width =3.1in]{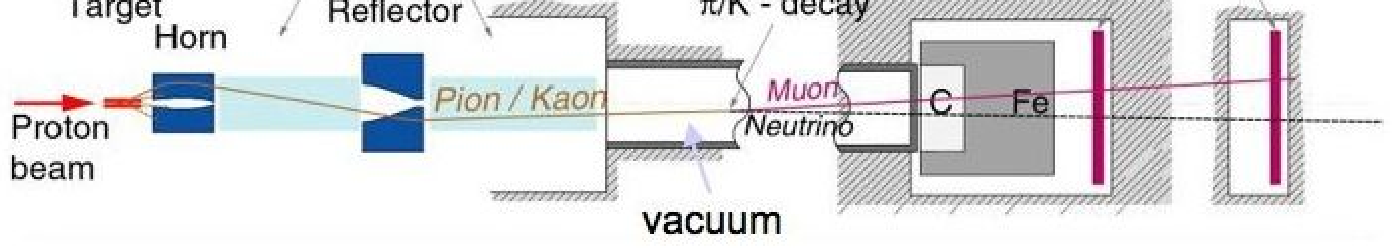}
\includegraphics[width =2.9in]{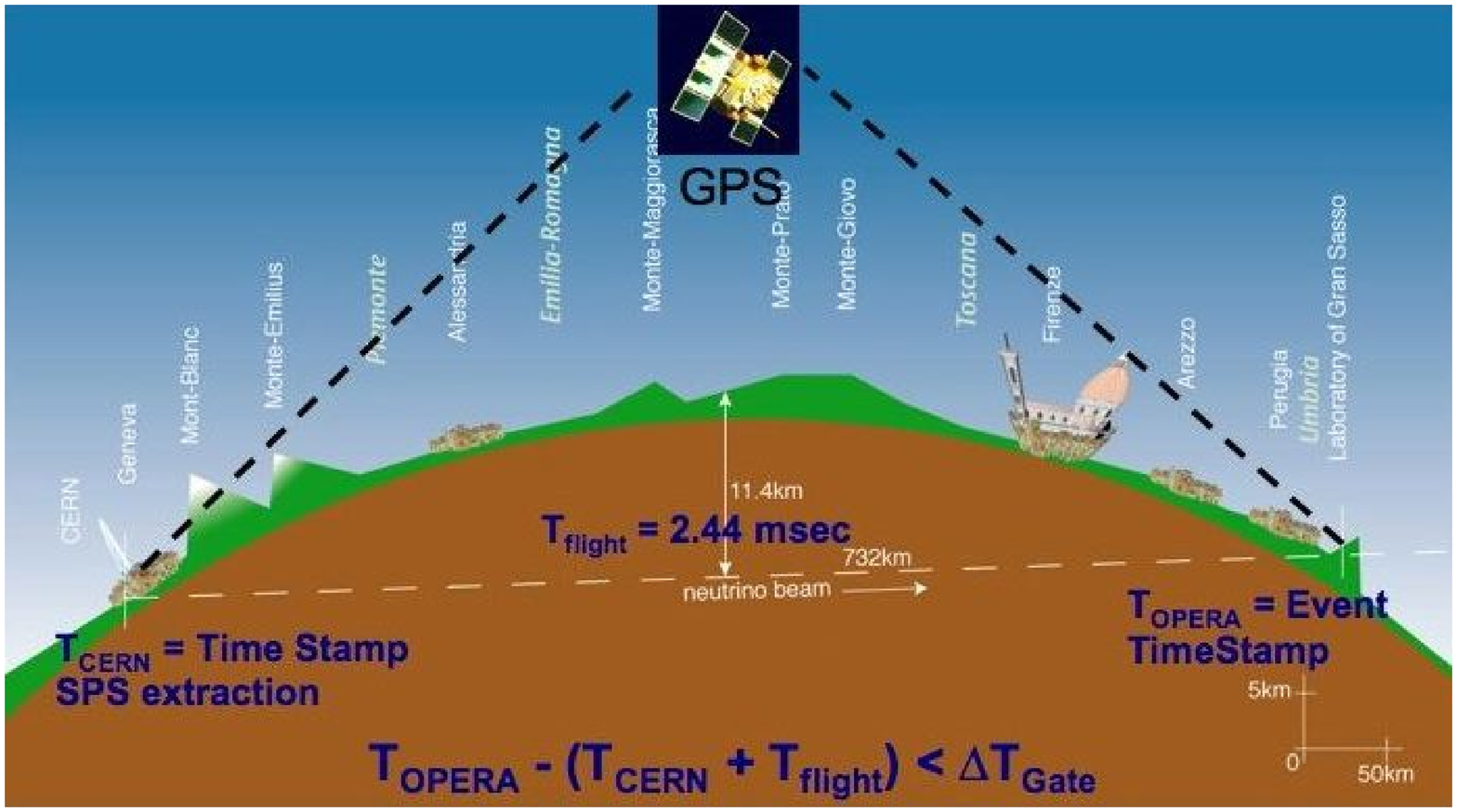}
\caption {\label{fig:CNGS}Left: The main components of the CNGS neutrino beam at CERN. 
Right: Sketch of the 730 km neutrino path from CERN to Gran Sasso and GPS time selection of events. }
\end{figure}

The {\bf CNGS $\nu_{\mu}$ beam}.
Fig. 4 left shows the main components of the $\nm$ beam 
at CERN \cite{CNGS}. A 400 GeV beam is extracted from the SPS and is 
transported to the target. Secondary pions and kaons are focused into a parallel beam by 2 magnetic lenses, called horn  and 
reflector. Pions and kaons decay 
into $\nu_{\mu}$  and $\mu$ in a decay pipe. The remaining hadrons are absorbed in the hadron stop. The $\mu$'s are monitored in 2 detectors.

Fig. 4 right shows the path of the 
CNGS $\nu_{\mu}$'s from CERN to GS. It also shows the synchronization via GPS of the atomic clocks at CERN and GS. The beam is optimised for producing 
a maximum number of CC $\nt$ interactions in OPERA. 
Fig. 5 left shows the underground layout of the SPS and of CNGS at CERN; Fig. 5 right shows the scheme of the SPS operation 
during a test run.
The mean $\nm$ energy is 17 GeV, the $\bar{\nm}$ 
contamination $\sim$2$\%$, the $\ne$ ($\bar{\ne}$) $< 1\%$ and the 
number of $\nt$ is negligible. The muon beam size at the 2$^{nd}$ muon detector at CERN is $\sigma \sim$1 m; this gives a $\nu_{\mu}$ beam size at GS of 
$\sigma \sim$1 km. The first low intensity test beam was sent to GS in August 2006 and 3 detectors (OPERA, LVD and Borexino) obtained their first events. 
The low intensity CNGS was stable 
and of high quality. The shared SPS beam sent a pulse of 2 neutrino bursts,
each of 10.5 
$\mu$sec duration, separated by 50 ms, every 12 s. 
A higher intensity beam expected 
for October 2006 did not happen because of a water leak at CERN. In September 2007 a 2$^{nd}$ test was successful, but a high intensity run was cancelled because of cooling problems.

\begin{figure}[!h]
\begin{center}
\includegraphics[width=4.8in]{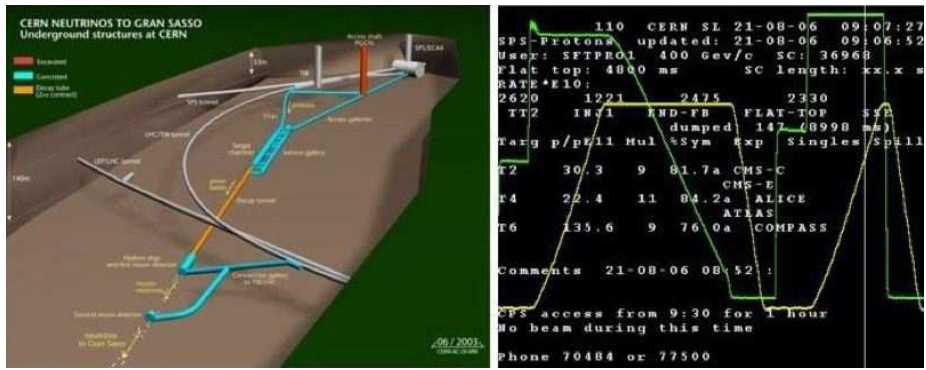}
\caption {\label{cngsbeam}Left: Underground layout of the SPS and of the CNGS beam at CERN. 
Right: Scheme of the SPS operation at CERN during the test runs.}
\end{center}
\end{figure}

The {\bf K2K} experiment confirmed the results of atmospheric $\nu_{\mu}$ oscillations \cite{k2k}. \par

The {\bf MINOS} experiment on the NuMi low energy neutrino beam is a large magnetised steel scintillator tracking calorimeter, complemented by a similar near detector and a calibration detector. The experiment obtained important results which confirm the atmospheric $\nu$ oscillation picture with maximal mixing and $\Delta m^2 = 2.38$ eV$^2$ \cite{minos}. \par

We shall now concentrate on the experiments at GS which may use the CNGS beam. \par

{\bf Borexino} is an electronic detector designed to study solar $\nu_e$'s, 
in particular those coming from Be$^7$ decays in the center of the sun \cite{http}. The important part of the detector is a large sphere, whose  central part was filled with liquid 
scintillator at the beginning of 2007. The 
outer sphere was started to be filled with water in 2006; during the first run only few meters of water were in the sphere and only few $\nu_{\mu}$ events 
were observed.   \par

{\bf LVD} is an array of liquid scintillators with a total mass of 
1000 t; it is designed to search and study $\bar \nu_e$'s from gravitational stellar collapses \cite{lvd}. LVD plans to be a neutrino flux monitor of the CNGS 
beam. LVD  has 3 identical ``towers'', 
each containing 8 active modules; a module has 8 counters of  
$1 \times 1 \times 1.5~ m^3$, filled with 1.2 t of liquid scintillator. CNGS $\nm$'s are observed through the detection of muons produced in neutrino CC interactions in the surrounding rock or in the detector and through the detection of the hadrons produced in neutrino 
NC/CC interactions inside the detector. In the 2006 test run LVD counted 50-100 muons per day and recorded $\sim$500 events \cite{lvd}.\par

{\bf OPERA} \cite{opera} is a hybrid-emulsion-electronic detector,
 designed to search 
for the $\nmnt$ oscillations in the 
parameter region indicated by the atmospheric neutrinos, confirmed
by the K2K and MINOS experiments.
 The $\nt$ appearance will be made by direct 
detection of the $\tau$ lepton, from $\nt$ CC interactions and the 
$\tau$ lepton decay products. 
To observe the decays, a spatial resolution of 
$\sim 1~\mu$m is necessary; this is obtained in emulsion 
sheets interspersed with thin lead target plates (Emulsion Cloud Chamber (ECC)). OPERA may 
also search for the subleading $\nmne$ oscillations and make a 
variety of observations using its electronic 
detectors. 
\begin{figure}[!h]
\begin{center}
\includegraphics[width=4.5in]{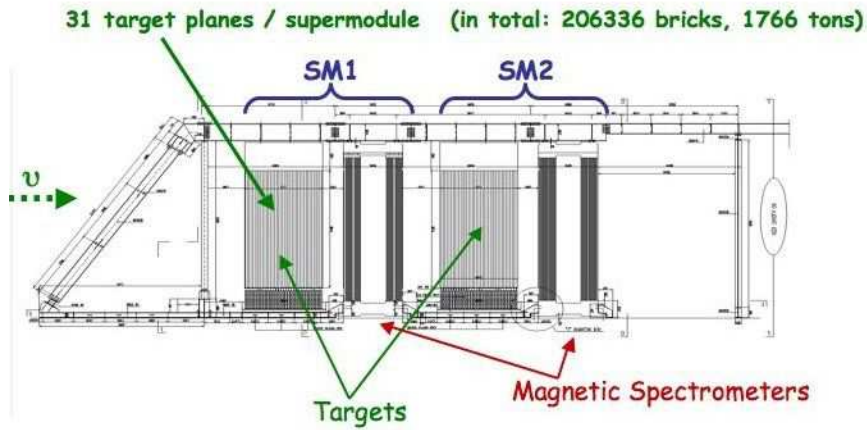}
\caption{\label{structure} Layout of the OPERA detector}
\end{center}
\end{figure}
The OPERA detector, Fig. \ref{structure}, is made of two 
identical supermodules, each consisting of a {\it target section} 
with 31 target planes of lead/emulsion-film modules (``bricks''), of a scintillator tracker detector and of a muon spectrometer. The final target mass is 1.55 kt. \par

The first electronic subdetector is an 
{\it anticoincidence wall} to separate $\mu$ events coming from 
interactions in OPERA from those in the material and rock  before OPERA.\par
The {\it target tracker} is made of scintillator strips, each 7 m long and of 25$ \times $15 mm$^2$ cross section. A wavelength shifting 
fibre of 1 mm diameter transmits the light signals to both ends. The readout 
is done by 1000 64 channel Hamamatsu PMTs. \par
The {\it muon spectrometer} consists of 2 iron magnets instrumented with 
{\it Resistive Plate Chambers} (RPC) and {\it drift tubes}. Each magnet is an 
$8 \times 8$ m$^2$ dipole with a field 
of 1.52 T in the upward direction on one side and in the downward direction on 
the other side. A magnet consists of 
twelve 5 cm thick iron slabs, alternated with RPC planes.\par
The {\it precision tracker} measures the muon track coordinates in the horizontal 
plane. It is made of drift tube planes, placed in front and behind each 
magnet and between the 2 magnets. 
The muon spectrometer has a $\Delta p / p \le 0.25$ for muon momenta $< 25$ GeV/c. 
 Two $45^{\circ}$ crossed planes of {\it glass 
RPC's (XPC's)} are installed in front of the magnets.\par
The {\it DAQ system} uses a 
Gigabit network of 1150 nodes. To match the data of the different 
subdetectors a ``time stamp'' is delivered by a clock using the GPS. Also the synchronization with the beam spill is done via 
GPS. 
The commissioning of each electronic detector was made with CR muons and with the CNGS at reduced intensity.

{\it Nuclear emulsions and their scanning.} The basic target module is a 
{\it ``brick''}, consisting 
of 56 lead plates (1 mm thick) and 57 emulsion 
layers. A brick has a size of $10.2 \times 12.7$ cm$^2$, a 
depth of 7.5 cm (10 radiation lengths) and a weight of 8.3 kg. Two 
additional emulsion sheets, the {\it changeable 
sheets} (CS), are glued on its downstream face. The bricks are arranged in 
walls. Within a brick, the achieved spatial resolution is $< 1~\mu$m and 
the angular resolution is $\sim$2 mrad.  Walls of target trackers provide the $\nu$ interaction trigger and the identification of the brick in which 
the interaction took place. \par
The bricks 
are made by the {\it Brick Assembling Machine} 
(BAM), which consists of robots for the mechanical packing of the bricks.
 The BAM, installed in the GS lab, produces 1 brick every $\sim$2 minutes. 
The bricks are handled by the {\it Brick 
Manipulator System} (BMS), made of two robots, each operating at one 
side of the detector.  

A fast automated scanning system with a scanning speed of 
$\sim20$ cm$^2$/h per emulsion ($44 ~\mu$m thick) is needed to cope with daily analyses 
of many emulsions. This is a factor of 10 increase with respect to past systems. For this purpose were 
developed the {\it European Scanning System} (ESS) in Europe \cite{ESS} and the 
{\it S-UTS} in Japan \cite{SUTS}. An emulsion 
is placed on a holder and the flatness is 
guaranteed by a vacuum system. 
By adjusting the focal plane of the objective, 16 tomographic images 
of each field of view are taken at equally spaced depths.
 The images are digitized, converted into a grey scale, sent to a vision processor and analyzed to recognize 
sequences of aligned grains. The 3-dimensional structure of a track 
in an emulsion layer ({\it microtrack}) is reconstructed by combining 
clusters belonging to images at different levels. Each microtrack 
pair is connected across a plastic base to form the {\it base track}. A set 
of connected base tracks forms a {\it volume track}.   \par

\begin{figure}[!h]
\includegraphics[width=3.8in, height=2.7in]{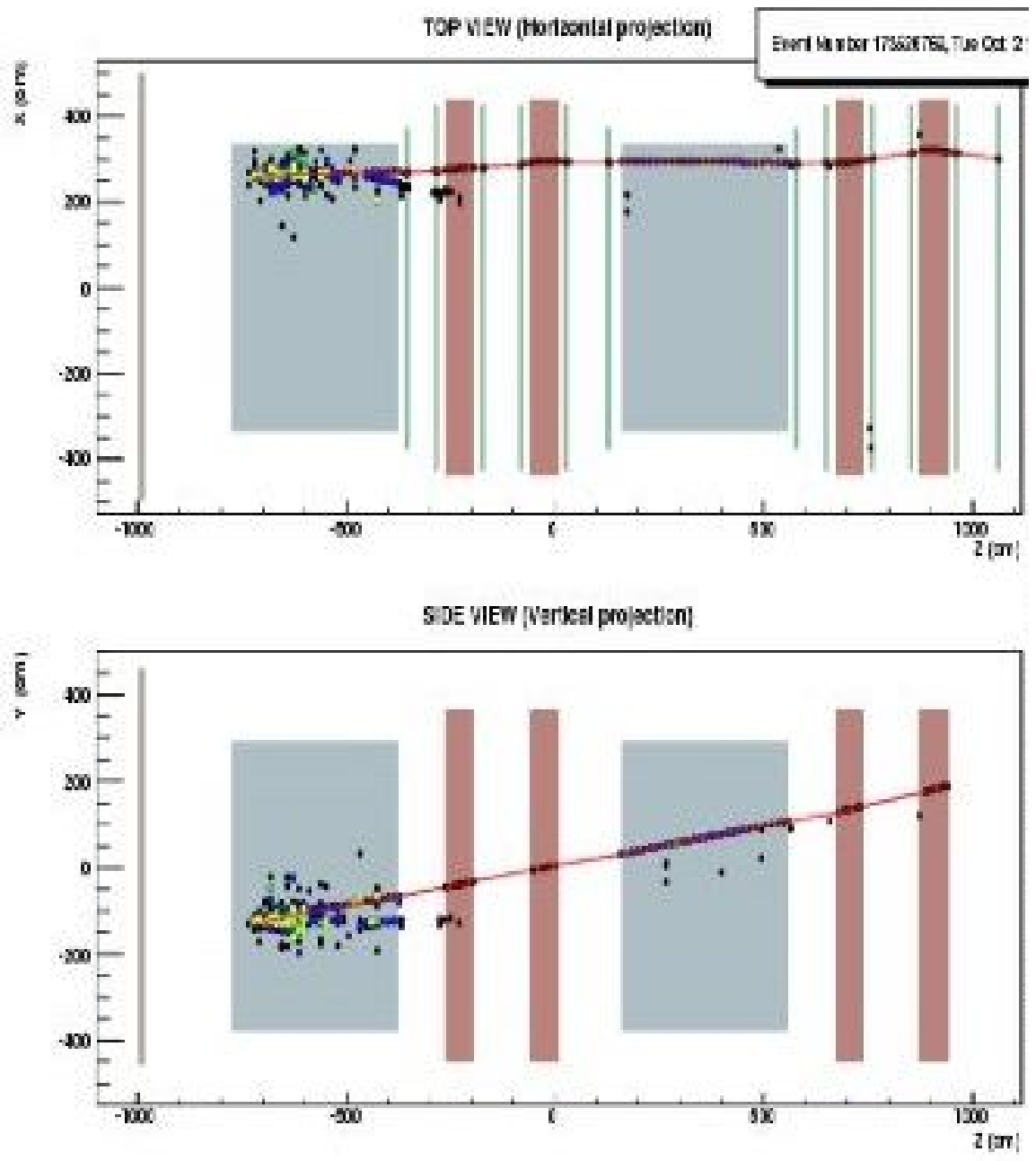}
\hspace{1pc}
\includegraphics[width=2.2in]{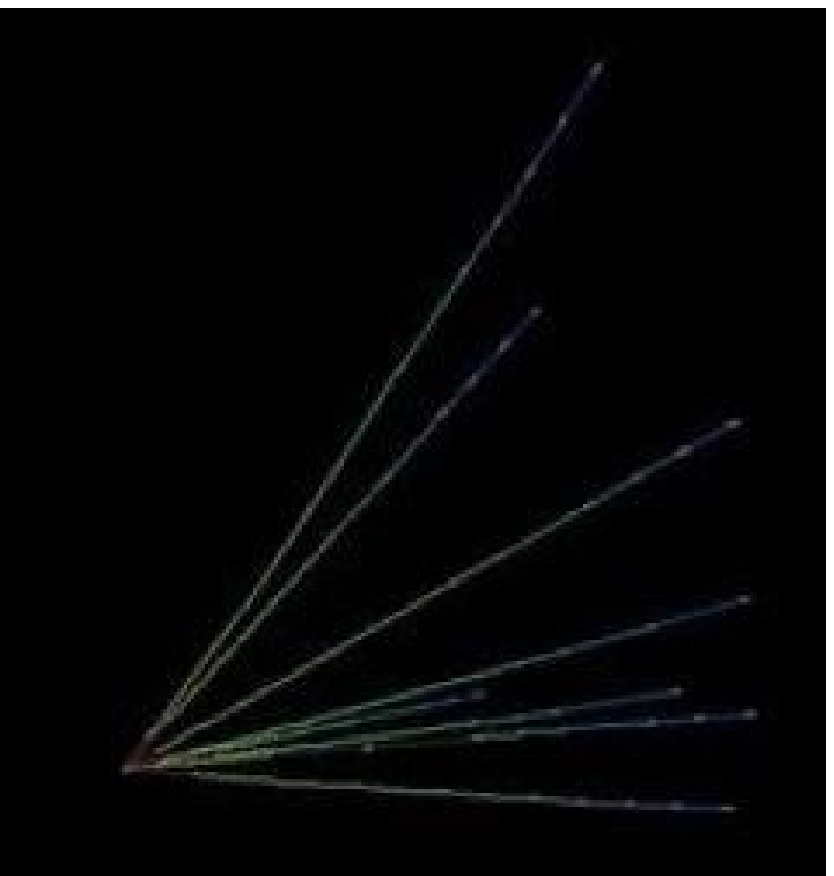}
\caption{\label{label} Left: display of one event with the $\nu$ interacting in a ``brick" in the first module, as seen by the electronic detectors; right: the interaction vertex seen in nuclear emulsions.}
\end{figure}

In the 2006 test run OPERA recorded 319 $\nu_{\mu}$ events, consistent with the 300 events expected from the delivered integrated intensity. A fit to the $\theta$ angle distribution 
of these reconstructed events on-time with the beam yielded a mean muon angle 
of 3.4$^\circ$ in agreement with the value of 3.3$^\circ$ expected for 
$\nu_{\mu}$ originating from CERN and travelling under the earth surface to 
the GS underground halls. During the same test run, a test of the CS procedure was performed, using an emulsion detector plane inserted in the SM2 target. 9 muons produced by 
neutrino 
interactions in the rock surrounding the detector crossed the CS plane surface. The test proved the capability of going from the 
centimetre scale of the electronic tracker to the micrometric 
resolution of nuclear emulsions \cite{opera}. During the short 2007 test run several interactions in the lead bricks positioned in the target section of SM 1 
($>$40\% of total mass) were detected by the electronic detectors and confirmed by tracks observed in the CS: the event vertexes were observed in the emulsions analysed by microscopes: Fig. \ref{label} left shows an interaction in the electronic detectors, Fig. \ref{label} right the vertex region in the emulsions. This test confirmed the validity of the methods to associate electronic detectors to nuclear emulsions.


\section{Conclusions}

The atmospheric neutrino anomaly became, in 1998, the atmospheric neutrino oscillation hypothesis with maximal mixing and $\Dm_{23} ~ \sim 2.4 \cdot 10^{-3}$ eV$^2$.\par
It was later confirmed with more data and by the first two long baseline experiments. All experiments agree on maximal mixing, while the $\Dm_{23}$ are:\\
Soudan-2 5.2   ,  MACRO 2.3  ,  SK 2.5  ,  K2K 2.7  , MINOS 2.38 $\cdot 10^{-3}$ eV$^2$. \\
These results come from disappearance experiments 
($\nu_{\mu}  \rightarrow  \nu_{\mu}$). An appearance experiment ($\nmnt$) with $\nu_{\tau}$ detection would solve conclusively the situation. \par

Other more exotic scenarios have been investigate like: Lorentz invariance violation, $\nu$ radiative decays, etc. Stringent limits were established for these hypotheses.

\vspace{0.2in}

{\bf Acknowledgments.} I acknowledge many colleagues at CERN, LNGS and Bologna. I thank drs. A. Casoni, M. Errico, M. Giorgini and R. Giacomelli for technical help.

\end{document}